\documentclass[showpacs,aps,superscriptaddress,prb,twocolumn,amsmath]{revtex4}
\usepackage{amssymb}
\usepackage{graphicx,subfigure,float}
\usepackage[percent]{overpic}
\usepackage{dcolumn}
\usepackage{bm}
\usepackage{color}
\usepackage{hyperref}
\begin{document}

\preprint{APS/123-QED}

\title{Large gap two dimensional topological insulators: bilayer triangular lattice TlM (M = N, P, As, Sb)}
\author{P. Zhou}
 \affiliation{Key Laboratory of Low-dimensional Materials and Application Technology, School of Material Sciences and Engineering, Xiangtan University, Xiangtan 411105, China}
\author{L. Xue}
 \affiliation{College of Physics and Optoelectronics, Taiyuan University of Technology, Taiyuan 030024, China}
\author{L. Z. Sun}
 \email{lzsun@xtu.edu.cn}
 \affiliation{Hunan Provincial Key laboratory of Thin Film Materials
and Devices, School of Material Sciences and Engineering, Xiangtan
University, Xiangtan 411105, China}
\date{\today}
\begin{abstract}
Based on density functional theory and Berry curvature calculations, we predict that \emph{p}-\emph{p} band inversion type quantum spin Hall effect (QSHE) can be realized in a series of two dimensional (2D) bilayer honeycomb TlM (M = N, P, As, Sb), which can be effectively equivalent to bilayer triangular lattice for low energy electrons.  Further topological analysis reveals that the band inversion between $p_{z}^{-}$ and $p_{x,y}$ of M atom contributes to the nontrivial topological nature of TlM. The band inversion is independent of spin-orbit coupling which is distinctive from conventional topological insulators (TIs). A tight binding model based on triangle lattice is constructed to describe the QSH states in the systems. Besides the interesting 2D triangular lattice $p-p$ type inversion for the topological mechanism, the maximal local band gap of the systems can reach 550 meV (TlSb), which provides a new choice for future room temperature quantum spin Hall Insulator (QSHI). Considering the advance of the technology of van der Waals passivation, combining with hexagonal materials, such as h-BN, TlMs show great potential for future 2D topological electronic devices.\\
\end{abstract}
 \maketitle
\section*{INTRODUCTION}
\indent Two dimensional topological insulator (TI), also known as quantum spin Hall insulator (QSHI), firstly theoretically proposed in graphene\cite{kane1,kane2}, draw more attention than three dimensional ones due to their only two direction helical edge states whose back-scattering is prohibited by time reversal symmetry (TRS). Such edge states are potentially applied in nano-electronic devices with dissipationless electron transport channels free from backscattering by any nonmagnetic impurities and defects, just behaving as a two-lane highway\cite{weng}. Nowadays, according to the types of lattice, the two dimensional TIs can be simply summarized as three categories: (i) traditional graphene-like hexagonal lattice materials. The representative examples are chemically functionalized single layer Sn\cite{Sntopo}, Ge\cite{Getopo}, Bi or Sb\cite{Bitopo}; (ii) square or rectangle lattice 2D materials including square-octahedral lattice of $MX_2$\cite{mx2_1,mx2_2,mx2_3,mx2_4}, buckled square lattice Bi$_4$F$_4$\cite{BiF}; (iii) the third major categories is triangle lattice systems. Au/GaAs(111) surface with $s-p$ band inversion and MXenes\cite{MXenes1,MXenes2} with $d-d$ band inversion are recently reported. According to energy band inversion in TIs the mechanism determined their topological properties, there are main four categories theoretically proposed. The most common one is \emph{s}-\emph{p} or \emph{p}-\emph{p} inversion. The HgTe/CdTe\cite{quanwell1}, InAs/GaSb\cite{quanwell2} quantum wells, and Bi$_2$Se$_3$ family compounds\cite{BiSe} belong to this category. The second one is the inversion between \emph{d} and \emph{p} bands, which is recently proposed in bismuth-based skutterudites\cite{dptopo}. Topological Kondo insulators\cite{kondo1,kondo2} recently predicted in SmB$_6$ AmN and PuTe family compounds belong to \emph{d}-\emph{f} type inversion. The \emph{d}-\emph{f} type inversion of topological Kondo insulators provide possibility to enhance the SOC energy gap of TIs. The last one is \emph{d}-\emph{d} inversion TIs. Transition-metal halide and oxygen functionalized MXenes\cite{MXenes1,MXenes2} recently proposed are typical examples for this classification. To find plentiful non-trivial 2D materials with the combination of lattices and inversion mechanisms is significantly important not only for understanding the ground physics of 2D quantum spin Hall effect (QSHE), but also for promoting exploration new materials in experiments.\\
\begin{figure}
\includegraphics[trim={0.0in 0.0in 0.0in 0.0in},clip,width=3.5in]{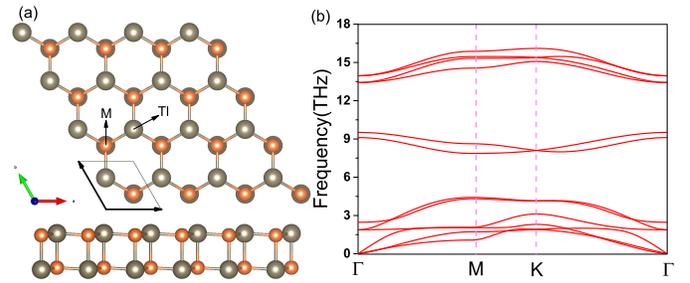}\\
\caption{(color online) (a) The side and top view of TlM bilayer. (b) Phonon band dispersion of TlN bilayer.}\label{fig1}
\end{figure}
\begin{figure*}
\includegraphics[trim={0.0in 0.0in 0.0in 0.0in},clip,width=5.0in]{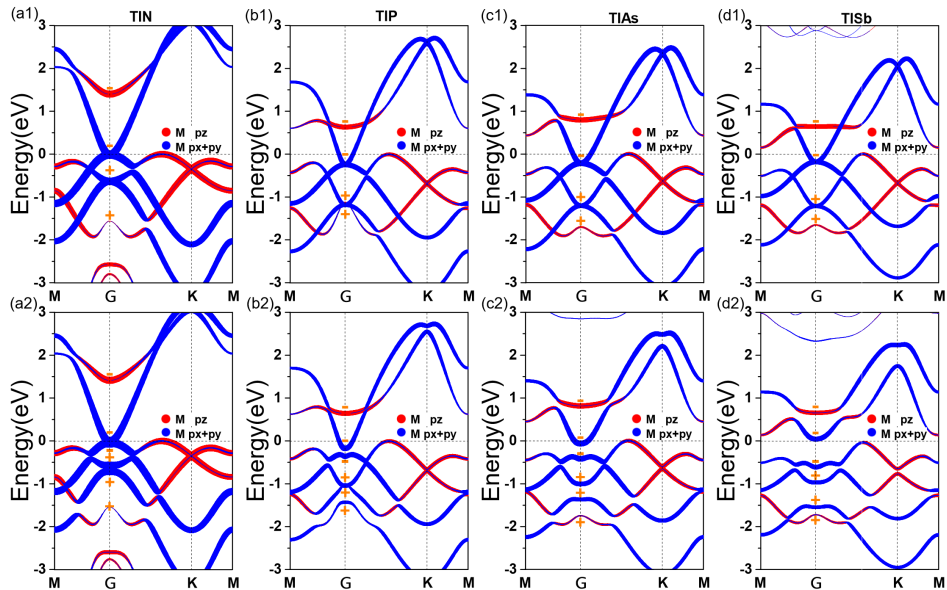}\\
\caption{(color online) Calculated PBE energy band of TlM without SOC (a1)-(d1) and with SOC (a2)-(d2). The signs in the figures represent the parity of wavefunction at $\Gamma$ point.}\label{fig2}
\end{figure*}
\indent Although large number of 2D TIs were theoretically predicated, unfortunately, the topological edge states only detected in few 2D materials, including HgTe/CdTe\cite{quanwell1} and InAs/GaSb\cite{quanwell2} quantum well, bilayer Bi\cite{Bi1,Bi2,Bi3,Bi110}, and 2D ZrTe$_5$\cite{ZrTe5_1,ZrTe5_2}. Moreover, the conditions to synthesis these 2D TIs are very rigours including extremely low temperature, ultrahigh vacuum, and high-accuracy molecular beam epitaxy growth. Thus, searching for feasibly experimental preparation 2D TIs is of great importance for promoting its applications in nanoelectronics. Furthermore, in the view point of applications, finding 2D TIs with large band gap is another critical issue to pave the way for QSHI in real nano-electronics devices. \\
\indent In present work, we predict a new family of QSH insulators based on bilayer hexagonal TlM (M=N, P, As, Sb). Ab initio molecular dynamics calculations prove that there are no obvious deformation in the bilayer crystal systems even the temperature up to 800 K. In combination with the phonon spectrum analysis, thermal and dynamic stability of the systems were confirmed. Further electronic investigation indicates that the special topological non-trivial nature of the systems comes from the band inversion between $p_{z}^{-}$ and $p_{x,y}$ of M atom with the help of equivalent to bilayer triangular lattice crystal field. Interestingly, such $p-p$ band inversion is independent of spin-orbit coupling (SOC) which is distinctive from conventional TIs, enriching the family of QSH insulators. Moreover, they are robust QSH insulators against external strain. The effect of SOC open an trivial bulk gap which can reaches 550 meV (TlSb) guaranteing their high temperature applications. Their almost flat hexagonal surface atom configuration makes it also appropriate to match with hexagonal boron nitride and easily forms topological heterojunction, which paves a new way for high temperature topological edge states detection in experiment.\\
\begin{table*}
\caption{Crystal constant ($a$), bond length in the plane (b$_{\parallel}$) and perpendicular to the plane (b$_{\perp}$), thickness (h), cohesive energy of each formula TlM (E$_h$), formation energy of bilayer TlM from two single layer TlMs ($\Delta$E), electron number of single M atom gain from Tl atoms with the Bader charge analysis (BC), local band gap around $\Gamma$ point (E$_{loc}$), and topological index (Z$_2$) of all the four systems.}\label{tab1}
\begin{tabular}{ccccccccc}
\hline
  structure      &  $a$(\AA)  &b$_{\parallel}$/b$_{\perp}$(\AA)& h (\AA) &E$_h$(eV/f.u.)& $\Delta$E(meV)& BC({$\vert$}e{$\vert$})& E$_{loc}$(meV) &  Z$_2$      \\
\hline
  TlN     & 3.977 &2.474/2.507& 2.507 & 4.79 & -370  &  0.82 & 54   & 1  \\
  TlP     & 4.691 &2.713/2.799& 2.799  &4.49 & -182  &  0.38 & 146  & 1   \\
  TlAs    & 4.862 &2.808/2.871& 2.871 &4.27 & -158  &  0.24 & 306  & 1 \\
  TlSb    & 5.195 &3.003/3.029& 3.029 &4.03 & -152  &  0.10 & 559  & 1   \\
  \hline
\end{tabular}
\end{table*}
\section*{COMPUTATIONAL METHODOLOGY}
\indent First-principles calculations were carried out with the Vienna ab initio simulation package (VASP)\cite{vasp1,vasp2}. The GGA of Perdew-Burke-Ernzerhof type exchange-correlation functional was used to simulate electronic interactions. OptB88-vdW exchange-correlation functional\cite{vdw1,vdw2} was applied to consider the Van der Waals interaction for BN-TlM heterojunctions. Spin-orbit coupling (SOC) was taken into account for topological properties computations. The cutoff energy for plane-wave expansion was 500 eV and in the self-consistent process $11 \times 11 \times 1$ k-point mesh was used. The crystal structures were fully relaxed until the residual force on each atom was less than 0.01 eV/{\AA}. A vacuum of 15 {\AA} was considered to minimize the interactions between the neighboring periodic images. Phonon spectrum\cite{phonopy} was employed to investigate dynamic stability of the systems. Considering the possible underestimation of the band gap within GGA, the hybrid density functionals(HSE06) functional\cite{HSE1,HSE2} was further applied to check the energy band characters. To explore the edge states, we cut the maximally localized Wannier functions (MLWFs) hamiltonian to a semi-infinite form. MLWFs were generated by using the wannier90 code\cite{wan90}.\\
\section*{RESULATS AND DISCUSSIONS}
\indent The bilayer TlM (M = N, P, As, Sb) was constructed by adopting the AB stacking of monolayer hexagonal TlM counterpart\cite{TlA}. The perpendicular bonding between the two monolayers results in the bilayer TlM (M = N, P, As, Sb) systems, the crystal structure is shown in Fig.\ref{fig1} (a). Different from the monolayer TlM, in bilayer TlM, the Tl and M atoms in the same sublattice (top-layer or bottom-layer) share the same plane. In the bilayer systems, each Tl (M) bonds with four nearest M(Tl) atoms, three of them locate in the same 2D plane with  Tl (M) and the rest one is located in the opposite plane. The three bonds in the plane are equal, and the bond perpendicular to the plane is a little longer than the bond in the plane. The bond lengthes of the TlM as listed in the Tab.\ref{tab1} indicate that the bond length of the four systems strongly matches with the atom radius of M. All the four systems are crystallographically subjected to the P-3M1 space group (No.164) with a point group D$_{3d}$. We listed the optimized crystal constant and thickness (namely, the spacing between two single layer TlM) in the Tab.\ref{tab1}. The results indicate that both the lattice constant and thickness of the bilayer systems show increasing trend from N to Sb in consistent with the variation of anion radius. To prove their feasible synthesis in experiments and stability, firstly, we calculate their cohesive energy derived from E$_{coh}$=(2E$_{Tl}$+2E$_M$-E$_{TlM}$)/4, the results are listed in the Tab.\ref{tab1}. The results indicate that in the energy stability respect the cohesive energies of bilayer TlMs are more stable than those experimentally synthesized 2D monolayer, such as black phosphorus (3.48 eV) and silicene (3.94 eV)\cite{silicene}. We also calculated the formation energy of the bilayer TlM from two single layer of TlM derived from ${\Delta}$E= E$_{bilayer}$-2E$_{monolayer}$, the results are listed in the Tab.\ref{tab1}. The negative formation energies indicate their easier experimental synthesis through the single layer TlM. To prove the dynamic stability of bilayer TlM, we calculated their phonon spectra. We take TlN as example and show its phonon spectrum in Fig.\ref{fig1} (b). The results show that imaginary frequencies can not be found in reciprocal space indicating their dynamical stability. The thermal stability of the bilayer TlM was further investigated by performing ab initio MD simulations. We adopted 3$\times$ 3 supercell in the simulations. In the AIMD simulations the NVT ensemble was used with a target temperature of 500 K and 800 K, maintained with a Nos\'{e}-Hoover chain thermostat state. The time step was set as 1 fs and the total simulation time is up to 6 ps. In the whole simulation period, the structure of TlN can be maintained, which indicates that it is thermodynamically stable.\\
\begin{figure*}
\includegraphics[trim={0.0in 0.0in 0.0in 0.0in},clip,width=5.0in]{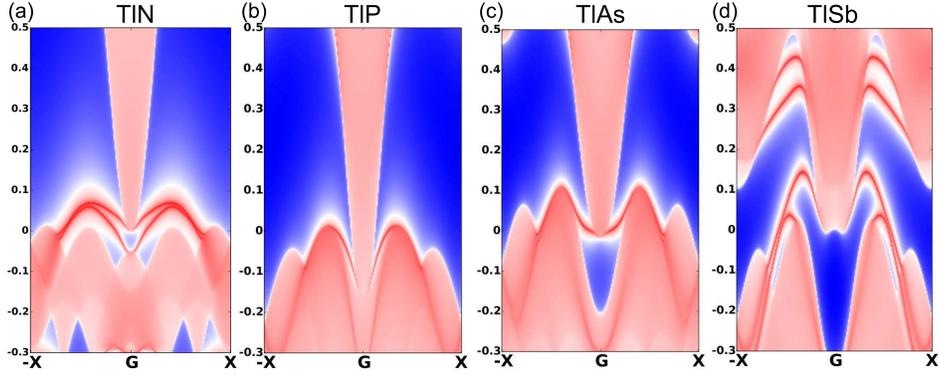}\\
\caption{(color online)  Energy and momentum-dependent local density of states for bilayer TlM on the zigzag edge. The Fermi level is set to zero. }\label{fig3}
\end{figure*}
\indent We then investigate the electronic structure information for these 2D materials. With the help of Bader charge analysis\cite{bader}, we find that all M atom obtain electrons from Tl atoms because of the different electronegativity between Tl and M (Tl is 1.62, N is 3.04, P is 2.19, and Sb is 2.05). The concrete values are present for all the four TlMs in Tab.\ref{tab1}. It gradually decreases with declining electronegativity of M atoms. The results indicate that the bonds of TlMs are all ionic covalent ones. The energy band structures of all the systems along high symmetry lines in reciprocal space without SOC ((a1)-(d1)) and with SOC ((a2)-(d2)) are shown in Fig.\ref{fig2}. When SOC is not turned on, the energy eigenvalues around Fermi level at the $\Gamma$ point for all the systems are degenerate with the two order representation of E$_u$. The partial band projections reveal that the low energy states mainly come form M sublattice atoms. Around $\Gamma$ point, M-p$_{x,y}$ state contribute to the degenerate electronic state around the Fermi level with negative parity, and M-p$_{z}$ states with positive and negative parity distribute on both sides of Fermi level with one order representation of A$_{1g}$ and A$_{2u}$, respectively. Unfortunately, the energy band extreme point along the high symmetry line $\Gamma$-K exceeds the degenerate energy at $\Gamma$  point with PBE, this is not advantageous to the detection of topological state in experiments. Our further calculations confirm that the states along $\Gamma$-K can be easily shifted to lower energy by applying small compress strain, which will be discussed below. When SOC is considered, a local band gap E$_{loc}$ is opened around $\Gamma$ for all the four cases and the values are listed in Tab.\ref{tab1}. The variation of them are match with the SOC strength from N to Sb, which is consistent with the discussions above that these electronic states are mainly come from M sublattice atom. To confirm the above PBE results, we also calculated the energy band with HSE06 and the results are shown in Figure S1 in the supplementary materials. Except shifting the extreme point under the degenerate point at $\Gamma$ for TlN, it have small impact on the low energy electronic states for other TlMs.\\
\indent To identify the topological nature of the four TlMs, we calculated the Z$_2$ number for all the four systems with the method of parity criteria proposed by Fu and Kane\cite{z2parity}. The parity eigenvalues $\delta$ of valence bands for all the four systems are as follows: $\delta_\Gamma$ = +1, $\delta_{M_{1,2,3}}$ = -1 (although metallization is happen for some structure, the energy bands around Fermi level are well separate especially at $\Gamma$ point, so the Z$_2$ number is well defined). Therefore, all the TlMs considered in present work are nontrivial topological materials with Z$_2$ = 1. The above results are obtained with SOC is turned on. However, its energy band inversion is independent of SOC which is distinctive from some conventional topological insulators (TIs), such as Bi$_2$Se$_3$\cite{BiSe}. The mechanism determined the non-trivial properties of the systems will be discussed below. The 2D nontrivial QSHI states in the TlM bilayer should support odd number of topologically protected gapless conducting edge states connecting the valence and conduction bands of the systems at certain k-points. To see these topological features explicitly, we perform the calculations of the edge states by cutting 2D monolayer into nanoribbon with the MLWF extracted from ab initio calculations. To eliminate the coupling between two edges, semi-infinite ribbons were used for all cases. The calculated results of all the systems based on PBE plus SOC are shown in Fig.\ref{fig3}. The results show that gapless edge states connect the valence states with the conduction states for all the four TlMs. The spin-projected LDOS\cite{zhounano} further prove that the edges state are all spin-polarized.\\
\begin{figure}
\includegraphics[trim={0.0in 0.0in 0.0in 0.0in},clip,width=3.5in]{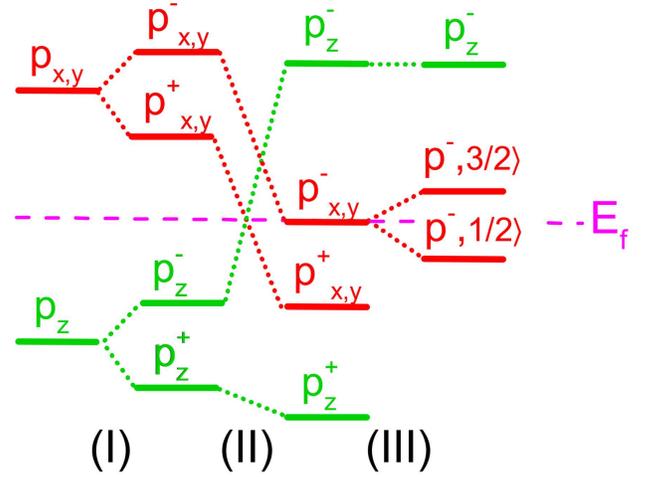}\\
\caption{(color online) Schematic of the evolution of energy levels at $\Gamma$ point for TlM 2D crystal. The stages (I), (II), and (III) represent the effects of turning on interface chemical bonding between the two single layers of TlM made up of the bilayer TlM, crystal field effect, and SOC in sequence, respectively.
  }\label{fig4}
\end{figure}
\indent Here we will discuss the origin the 2D topological states. According to previous report that the band inversion determined the topological property of Bi$_2$Se$_3$ is induced by strong SOC of Bi atoms\cite{BiSe}. However, in our case, the band inversion comes from the effect of crystal field and the SOC open an trivial bulk gap, just like previous proposed 2D QSHI tin films\cite{Sntopo} and functionalized germanene\cite{GeI}. To illustrate the band inversion process explicitly, we plot the p$_z$ projected energy band in Figure S2 in the supplementary materials with increasing thickness of TlN. In this process, the in-plane crystal constant is fixed. It is clearly shown that the reversal band order start recover to normal order for p$_z$ after the thickness added 1.2 \AA. Based on the above results and partial band projections in Fig.\ref{fig2}, the schematic of energy level sequence around $\Gamma$ corresponding to three stages of turning on the interface resonant coupling, crystal field, and SOC is shown in Fig.\ref{fig4}. Firstly, when the two TlM single layers are totally isolated, the energy position of p$_z$ behaved as bonding state is lower than that of anti-bonding p$_{x,y}$ as shown in Fig.\ref{fig4} due to the chemical bonding in each single layers. As for stage (I), the interaction in the plane of monolayer form bonding p$^{+}_{xy}$ and anti-bonding states p$^{-}_{xy}$, when the two layers close to each other and their thickness is still larger than h+1.2\AA, the p$_z$-orbit in top and bottom planes resonant coupling with each other to form bonding p$^{+}_z$ and anti-bonding states p$^{-}_z$. When crystal field effect is taken into account in stage (II), namely M atom chemically bonds with Tl atom in the opposite layer and form bilayer trigonal lattice, band inversion occurs between p$^{-}_z$ and all p$_{xy}$ states and leading to a nontrivial band order in bilayer TlM. For the inverted bands, the states p$^{-}_z$ and p$^{+}_{xy}$ are occupied and the Fermi level locate on or little lower than the degenerated p$^{-}_{xy}$ states, thereby the systems behave as gapless semiconductor (TlN) or metals (TlP, TlAs, and TlSb). After further introducing SOC in stage (III), the degeneracy of p$^{-}_{xy}$ is lifted and an local energy gap is opened. During this stage, the inverted band order remains unchanged without parity exchanging between occupied and unoccupied states. Consequently, the band inversion in bilayer TlM 2D crystal stems from crystal field effect instead of SOC. It is interesting to note that the p$_z$-p$_{x,y}$ band inversion in the 2D bilayer TlM crystal is a rare kind of energy inversion for QSHI.\\
\indent From the atom-projected energy band around Fermi level, we know that the low energy electronic states around $\Gamma$ mainly come from M sublattice. In such view of point, the M atoms form bilayer triangle lattice intermediated by Tl. In order to capture the mechanism of QSHE in bilayer TlMs, we construct a Slater-Koster (SK) type\cite{kstb} tight binding model with a minimal basis (p$_x$,p$_y$,p$^{-}_z$) of M atoms under triangle lattice as below. Due to the almost flat hexagonal lattice for bilayer TlMs, just like 2D graphene, we can ignore the interaction between p$_{x,y}$ and p$_z$ states.
\begin{gather*}
H = H_0 + H_{SOC}\\
\end{gather*}
with
\begin{gather*}
H_0 = \begin{pmatrix}h_{11} & h_{12} & h_{13} \\ h^{*}_{12} & h_{22} & h_{23}\\ h^{*}_{13} & h^{*}_{23} & h_{33} \end{pmatrix};\quad H_{SOC} = \begin{pmatrix}0 & -\lambda \emph{\textbf{i}} & 0 \\ \lambda \emph{\textbf{i}} & 0 &0\\ 0 & 0 & 0 \end{pmatrix}\quad
\end{gather*}
\begin{align*}
& h_{11} = \varepsilon_{xy} + (3t_{pp\sigma}+t_{pp\pi})cos(\frac{\sqrt{3}}{2}k_x)cos(\frac{k_y}{2})\\
& +2t_{pp\pi}cos(k_y)\\
& h_{12} = (t_{pp\pi}-t_{pp\sigma})\sqrt{3}sin(\frac{k_y}{2})sin(\frac{\sqrt{3}}{2}k_x)\\
& h_{22} = \varepsilon_{xy} +(3t_{pp\pi}+t_{pp\sigma})cos(\frac{\sqrt{3}}{2}k_x)cos(\frac{k_y}{2})\\
& +2t_{pp\sigma}cos(k_y)\\
& h_{33} = \varepsilon_{z} + 2t^{*}_{pp\pi}(2cos(\frac{\sqrt{3}}{2}k_x)cos(\frac{k_y}{2})+cos(k_y))\\
& h_{13} = h_{23} = 0
\end{align*}
\indent where $\varepsilon_{z}$ and $\varepsilon_{xy}$ are onsite energies for p$^{-}_z$, p$_{xy}$ orbits, respectively. The term t$^{*}_{pp\pi}$ is the next nearest (NN) hopping parameter between p$_z$-orbitals. $t_{pp\sigma}$ and $t_{pp\pi}$ are NN hopping parameters corresponding to the $\sigma$ and $\pi$ bonds formed by p$_{x,y}$ orbitals. By diagonalizing the Hamilton matrix at $\Gamma$ point, we obtain the three eigenvalues with the values of $ E_{p^{-}_z} = \varepsilon_{z} + 6t^{*}_{pp\pi}$ and $E_{p^{\pm}_{xy}} = \varepsilon_{xy} + 6t_{pp\pi} + 6t_{pp\sigma} \pm \lambda$. It is indicated that the SOC does not affect the $E_{p^{-}_z}$. If $\lambda$ = 0, the energy bands of p$_{xy}$ at $\Gamma$ are degenerate, and finite $\lambda$ would open a local gap with the value of 2$\lambda$.  We plot the low energy bands around $\Gamma$ point in Fig.5(a) and (b) with same parameters except SOC parameter $\lambda$. Without SOC, the electronic state of p$_{x,y}$ intersect each other with quadratic band touching. When including SOC, even an infinitesimal value of $\lambda$ would removes the quadratic band touching between the two bands, turning the system into a QSH phase.\\
\begin{figure}
\includegraphics[trim={0.0in 0.0in 0.0in 0.0in},clip,width=3.5in]{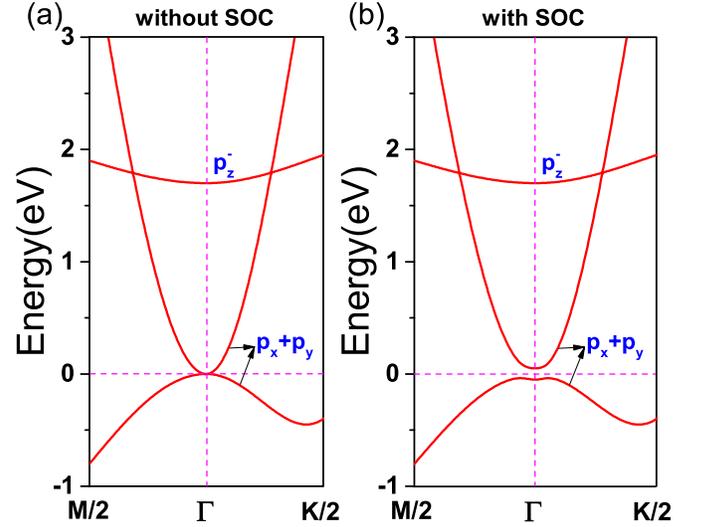}\\
\caption{(color online) The energy band structure from tight-binding model with the parameter $\varepsilon_{xy} = 2.4 eV, \varepsilon_{z} = 2.0 eV, t_{pp\sigma} = -1.6 eV, t_{pp\pi} = 0.8 eV, t^{*}_{pp\pi} = -0.05 eV$, (a) $\lambda = 0.0$ and (b) $\lambda = 0.05$. }\label{fig5}
\end{figure}
\indent Here we will discuss how to shift down the energy band extreme point around the Fermi level along $\Gamma$-K producing global band gap in the systems. Massive theoretical and experimental reports indicate that strain is an effective way to modulate electronic states of 2D materials\cite{strain1,strain2,strain3,strain4}. In present work, we use external strain to lower the energy band extreme point around the Fermi level along $\Gamma$-K of the systems. We take TlN as an example to study the effect of biaxial strain on its electronic structure. The band structures of TlN in function of biaxial external strains are shown in Figure. S3 in the supplementary material. The results indicate that when the compressed external strain reaches up to -3\% the extreme point along $\Gamma$-K shift below the degenerate energy point at $\Gamma$. Considering the SOC effect, the external strain will open a global band gap which is important for the systems to apply in nano-electronic devices. However, the expanded strain almost unaffect the electronic states of TlN. To test the multilayer stacking effect on the topological properties of the systems, we calculated the band structures of the systems with ABA (ABAB) stacking three (four) layers hexagonal TlN without and with SOC. The results of TlN are shown in Figure S4 in the supplementary materials as example due to the similar characteristics for the four systems. The results show that the multilayer stacking systems share similar band structures for low energy electronic states with that bilayer case. Z$_2$ calculations also confirm their nontrivial topological properties for the multilayer stacking systems.\\
\begin{figure}
\includegraphics[trim={0.0in 0.0in 0.0in 0.0in},clip,width=3.5in]{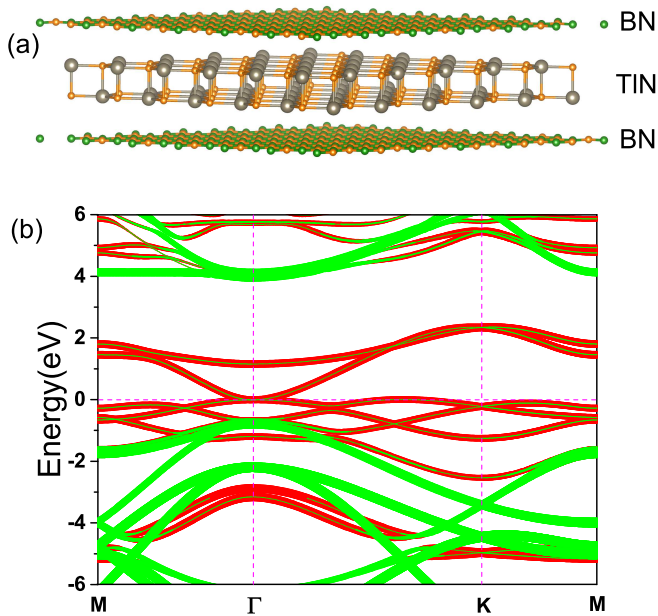}\\
\caption{(color online) The model for TlN sandwitched by h-BN (a) and its band structures (b). Red and green lines in the band structures represent the projection of all TlN atoms and all h-BN atoms, respectively.}\label{fig6}
\end{figure}
\indent Finally, let us discuss how to synthetize the bilayer TlMs and reserve their topological properties in experiments. We know that monolayer phosphorene attracts much attention due to its high carrier mobility and direct moderate band gap. However, it undergoes rapid degradation when it is exposed to ambient environment hampering its application. The TlM also show such problem due to its surface is easily contaminated by external chemical group. Taking TlN as an example, we find that H atom can easily adsorb on its surface, and the adsorption energy are 0.97 eV/H and 3.27 eV/H for Tl and N top sites, respectively. The band structures of H decorated TlN without and with SOC are shown in Figure S5. Without SOC, the degenerate energy level still remains around $\Gamma$ for Tl-H system, but the N-H system transforms to pure metal. The topological index of the two systems with SOC shows that when H adsorbed on Tl atom is still non-trivial, however, the topological nature disappears for H adsorb on N atom. This phenomenon is consistent with the origin (N sublattice) of its topological properties. Therefore, to reserve the novel properties of the 2D materials, surface protection is crucial. Van der Waals passivation is an effective approach for protecting 2D material from surface contamination as recently reported\cite{Pexp}. To make full use of the non-trivial properties of TlM, we construct the sandwich structure that the top and bottom surfaces of TlM are van der Waals passivated with single layer h-BN, the diagram of the model is shown in Fig.6(a). The h-BN can eliminate the influences of substrate and surface adsorption on the p$_z$ states of the systems. The partial projection energy band, taking TlN as example in Fig.6(b), clearly reveals that the low energy electronic states of TlM can be effectively reserved in this composite structure. Such model will be an effective apporach for reserving the non-trivial properties of 2D materials, especially for the systems whose topological properteis relate to outplane states.\\
\section*{CONCLUSION}
\indent In present work, first-principles density-functional calculations have been performed to study the electronic and topological properties of a series of TlM(M = N, P, As, Sb). All the bilayer systems studied are 2D TIs with nontrivial $Z_2$-type topological invariants. These 2D TIs have sizable bulk local band gaps ranging from 54 to 559 meV, which ensure their QSH effect at room temperature. Interestingly, the topological nature derives from the $p$-$p$ bands inversion of M sublattice. Considering recent experimental success in producing van der Waals passivation, all the TlMs show great potential for future 2D topological electronic devices. We are looking forward to the experimental confirmation of our prediction.\\
\begin{acknowledgments}
This work is supported by the National Natural Science Foundation of China (Grant No. 11574260 and 11547213), the Qualified Personnel Foundation of Taiyuan University of Technology(QPFT) (Grant No. tyut-rc201433a) and scientific research innovation project of Hunan Province(CX2015B219).
\end{acknowledgments}



\end{document}